# The Even Sheen of AI: Kitsch, LLMs, and Homogeneity

**Gyburg Uhlmann**

University of Technology Nuremberg
Nuremberg
Germany
gyburg.uhlmann@utn.de

## Abstract
The exploding use and impact of Chatbots such as ChatGPT that are based on Large Language Models urgently call for a language which is fit to clearly describe functions and problems of the production process and qualities of the Chatbots' textual and image output. Recently, the discussion about appropriate and illuminating metaphors to describe LLMs has gained momentum. As an alternative to well-established metaphors such as 'hallucinating' and 'bullshit', we propose 'kitsch' as a new metaphor. As an internationally widespread term from literary and cultural studies, we argue that 'kitsch' is particularly suitable for analytically illuminating a previously neglected feature of LLM-based images and texts: their tendency to produce homogeneous and average content, which is becoming increasingly dominant as the proportion of AI-generated content on the internet grows. This is leading to the equalisation of language, style and argument. In view of the potential negative consequences of this averaging, including for human content producers on the internet, we advocate combining methods and insights from kitsch studies with AI research, philosophy, and communication studies in order to better understand the phenomenon and develop countermeasures.

## Keywords
Large Language Models (LLMs), homogeneity, metaphors for AI, ethical implications, plausibility, rhetoric

## 1. Introduction
Since 2022, it has become increasingly popular to call the output of LLMs bullshit based on the concept introduced by philosopher Harry Frankfurt (On bullshit, Princeton 2005). The application of this concept has become fashionable here, like in many other areas such as political communication or ethical manipulation studies. Some scholars argue in favour of a paradigm shift away from the metaphor of hallucination or confabulation, which has been used to describe both unwanted failures and fabricated outputs (Hicks et al., 2024). To call LLMs bullshitters addresses — as does the metaphor "stochastic parrots" (Bender et al., 2021) — the very method by which LLMs produce their output and characterizes it in general and overall as bullshit, i.e., as something that was produced without any concern for truth (Gunkel & Coghlan, 2025; Fisher, 2025).

All these terms, thus, target factual misalignment or indifference to truth, but they overlook a persistent property of LLMs: a structural tendency toward homogeneity or averageness and form-level plausibility (Rudko & Bonab, 2025; Agarwal et al., 2024[1]). The homogeneity and averageness can have negative effects on the diversity and individual innovativeness of text production (Doshi & Hauser, 2025) not only of LLMs but also of human producers and recipients (Lee et al., 2025). They can also cause issues regarding the faithfulness of LLMs, resulting in reduced trustworthiness (Agarwal et al., 2024).

We will demonstrate why exactly this is a problem (in terms of communication, recognition, opinion forming and creativity) and why calling this bullshit (Rudko & Bonab, 2025) is not helpful.

As an alternative, we propose 'kitsch' as a new metaphor that combines various positive and negative characteristics. In particular, it is capable of illuminating the emotional and persuasive effects of typical LLM-generated texts or images and the production and reception conditions that result in homogeneous texts or images.

The definition of 'kitsch' has long been and is still debated in literary, art and cultural studies. We start with the following hypothesis: In 'kitsch' the author or artist seeks to superficially intensify emotional experience and perception by means of conventional language and the accumulation of established linguistic images and patterns. 'Kitsch' is a mass-produced item that does not develop its vividness, stylistic quality or plausibility from the plot or subject matter, but rather derives these qualities from high-quality linguistic or artistic products, applying them uniformly to make the item accessible and to create pleasurable emotional experiences, such as being moved (Genz, 2011; Braungart, 2002).

Thus, adopting the kitsch metaphor shifts the focus from truth failure to the averaging and smoothing of LLMs as core features of their production. This helps to explain why 'good-looking' texts and images can be persuasive without providing understanding and the negative impact this can have (Doshi & Hauser, 2025).

Because these features are essential to the generation process of content by LLMs, external aid and reflection are required in order to counterbalance the negative impact of the homogeneity of LLMs on ethical and social behaviour, political communication and human text production and reception.

This aid can be found in debates about kitsch in literary studies. We will demonstrate that the origins of this complex metaphor lie in the long history of the development and differentiation of the concept of kitsch as an antidote to art in literary studies. We will advocate a crossover from literary studies to information technology, revealing future perspectives beyond the concept of kitsch for describing new technologies. This will inform policymakers and the public adequately, while highlighting potential benefits and transformative risks (Hicks et al., 2024).

The paper, thus, makes four contributions: (1) We define kitsch-like output and distinguish it from "hallucination" (perceptual failure metaphor) and "bullshit" (speaker-intention). (2) We derive a set of stylistic markers (use of patterns, emotional address, glossy elevation) from classical and modern rhetoric and literary theory and show how they map onto LLM generation. (3) We argue that these markers persist under retrieval-augmented prompting and "truthful" instructions, indicating that they are essential to currently existing LLMs. (4) We outline countermeasures beyond factuality checks: fostering stylistic diversity, tethering plausibility to content, and redesigning evaluation criteria.

---

[1] Agarwal et al. (2024, 1): "We highlight that the current trend towards increasing the plausibility of explanations, primarily driven by the demand for user-friendly interfaces, may come at the cost of diminishing their faithfulness".

The structure of the paper is as follows: Section 2 motivates the homogeneity claim; Section 3 contrasts the prevailing metaphors with our proposal; Section 4 develops the kitsch framework and markers from rhetoric and literary studies; Section 5-6 discuss limitations, interdisciplinary responses, implications and future perspectives; with Section 7 we conclude with some remarks on the necessity of interdisciplinary research in the field.

## 2. Why do LLMs like ChatGPT produce homogeneous texts?

Programs such as ChatGPT, or rather their underlying language models[2] which use machine learning algorithms based on neural networks, learn from hundreds of billions of words in — at least today mostly — human-generated texts by a self-supervised learning method which words are likely to follow each other, and output the most probable word in each context. These programs use a large amount of sample data to calculate the probability of how human texts and reasoning processes on certain topics and text forms are normally structured on average (Agarwal et al., 2024). LLMs learn the patterns of word order and relationships between words or propositions in natural language and thus mimic the average form of texts that humans produce for a specific topic or task and by that produce more and more homogeneous texts and images (Endacott & Leonardo, 2024). This means that they are prone to manipulating the form of language (Shah & Bender, 2022) and producing probable and plausible content, including fabricated information calculated to appear genuine. This kind of mimesis is not merely mirroring or echoing its sources (Bender et a., 2021), but is obviously creative in that it builds abstractions, recombines data, and sets up new orders (Arkoudas, 2023) — all on the basis of the learned patterns and probabilities. The pursuit of statistically calculated probability and plausibility results in average output and creates a sense of positive familiarity in the human recipients and a readiness to show misplaced trust and over-reliance (Agarwal et al., 2024; Heersmink et al., 2024).

The averageness and homogeneity are essential, indispensable and irreducible properties, whereas deviating from the facts and inventing non-existent sources are not. Thus, while deviation from facts can be reduced or eliminated by two LLMs checking each other's results, referring the prompt question to a non-generative system, or using Retrieval Augmented Generation (RAG), where searches are launched in external sources (see Guu et al., 2020; Lewis et al., 2021; Yang & Fujita, 2024), the averageness of LLMs cannot be eliminated. This is because it is integral to how they function. In fact, if two LLMs attempt to eliminate homogeneity or check other sources, they will actually exacerbate it.

This process is enforced and accelerated if the training data contains AI-generated texts: When a model is trained over several generations using AI-generated texts, homogeneity of the texts increases until the model collapses and produces more and more repetitions (Shumailov et al., 2024). Rudko and Bonab (2025) rightly emphasize here that it is not enough to punish word repetitions as a countermeasure, because this is a structural problem of structural averaging (Wenger, 2024). As a result of this inbreeding training, the models 'forget' the less frequent elements in the original human-generated training data (Wenger, 2024), they eliminate them in order to adopt a more flat, polished style. Since the amount of machine-generated content will increase, so will the problem of homogeneous, average, or mediocre texts. Without countermeasures, all the features that make our texts unique and characteristic will gradually disappear from the web (Lee & Jeon, 2025).

---

[2] In the following we will synonymously speak of chatbots that are based on LLMs and LLMs. We will also refer to the products of these by using "AI generated texts or images" or "texts or images produced by LLMs". For our purposes the differentiation between the chatbots that are based on LLMs and the LLMs themselves is not necessary.

An important factor in this process is the pursuit of plausibility by the algorithms. But how can we define plausibility and probability in the context of our questions for human users? Due to the linguistic and rhetorical nature of the sought-after probability, rhetoric studies come into play (McKee & Porter, 2020[3]). Probability is generally understood as a measure of expectation for an uncertain event.[4] In rhetoric, probability forms the basis of persuasion, particularly in situations where absolute proof is unavailable. It connects the speaker, the message and the audience by providing premises that are both acceptable and understandable to listeners. In this context, plausibility refers to subjective probability with regard to comprehensibility and credibility. Something is credible if it can be convincing to a wide range of potential opinions and statements.

As AI texts are programmed to consider factors that make them appear plausible to human users, such as transparency regarding sources and limitations in knowledge, people are even more inclined to trust them (Zhou et al., 2023). AI texts therefore appear familiar and plausible because they are programmed to find the most probable sequence of words and propositions. However, the factors that lead to this trust are largely unspecific to the individual text produced. In other words, it is not the specific content of a text that seems plausible, but rather the way in which these texts are generated. Consequently, the attained plausibility does not allow any conclusions on the individual truth value of the single propositions, because it is only an abstract superficial feature of the generated text.

Here we introduce a distinction which comes from rhetoric studies, between a good rhetoric and a bad or sophistic rhetoric. We define good rhetoric, following Aristotle (Uhlmann, 2019), as a technique that deals with finding out what is convincing for a specific audience in a particular situation and presenting it in a plausible way. In contrast, bad or sophistic rhetoric is a technique whose goal is to persuade listeners to accept any opinion held by the speaker, or in the traditional phrase "to make the weaker argument stronger" (Diels & Kranz 2005, Protagoras, fragment A 21 (= Aristotle, *Rhetoric* 1402a23); Guthrie, 1969). In a bad rhetoric, the technique of persuasion is abstract from the specific position or situation. It mainly uses general methods and patterns in order to persuade or manipulate the listeners. Such a bad rhetoric is also used in kitsch literature and, as we will see, is applied in LLMs and their plausibility algorithms.

This abstract plausibility has several disadvantages. First and foremost, human users cannot conclude from the text's plausibility whether the individual text is probable or true. Plausibility remains an added-on surface quality, and remains abstract from the individual context.[5]

Furthermore, texts produced by ChatGPT and other chatbots are (mostly) linguistically flawless and smooth, and typically of a high quality, usually even of a stylistically higher quality than many human-made products on the Web 2.0. (Doshi & Hauser, 2024; Heersmink et al., 2024; Sastre et al., 2024). They thus generate trust and admiration, particularly when they intensively imitate the characteristics of good texts and eliminate flaws and underperformance. We do not need to apply a Romanticist concept, as Rudko and Bonab (2025) implicitly did, to explain why AI-generated texts lack something that human-generated texts have. Without necessity, we should not presuppose a quality that cannot be grasped rationally like the French "Je ne sais quoi" or in German continental modern aesthetic the *Geschmack* (taste) and attribute it to human-made texts. Instead, we should acknowledge that human-written texts *may* be better than average because of the specificity of their analysis and description of a given context, but they certainly do not always achieve this, and are often worse. AI-generated texts also reduce the stylistic diversity of text output by bringing these heights and depths down to the same intermediate average level (Padmakumar & He, 2024). On average, the stylistic quality of texts that look like a glossy brochure is likely to be perceived and received more positively than the more diverging sample of human-generated texts because of their homogeneity.

---

[3] McKee & Porter (2020) emphasize the importance of taking into account rhetorical qualities in the interaction between humans and between humans and machines.
[4] For a full account of philosophical theories and concepts see Eagle (2010). In our context we only need a common-sense approach that proves useful for rhetorical purposes. See Schuessler (2023).
[5] Another issue is that by striving for abstract plausibility, the system may fail to provide faithful explanations of how and what it is doing (Agarwal et al., 2024).

# 3. Bullshit and Hallucination as Descriptions for LLM Text Production (Problems)

Before we consider the advantages and power of the metaphors used so far to explain the operating mode and performance of chatbots based on LLMs, we will define what a metaphor is and how it works.

A metaphor is a linguistic transfer in which an expression from one area of meaning is transferred to another without a comparative word ("like," "as") in order to illustrate or reinterpret it. It is based on the recognition of a similarity and structural correspondence (the "tertium") between the two areas (Eggs, 2001). Part of the appeal of the metaphor, and a prerequisite for its heuristic value, is that it links and compares two very different contexts via the bridge of the *tertium*. In this interpretation, the structural relationship between the two areas must be deciphered, thereby elucidating the target object of the comparison. When using a metaphor, both the producer and the recipient are expected to stick to the third element of comparison, since metaphors are not designed to be completely parallel and similar in every respect.

When we understand "hallucinations" or "bullshit" as metaphors for LLMs, it is important to identify the *tertium*, i.e., the point of comparison (Mitchell, 2024; Nehrlich, 2024)[6].

Hallucinating means perceiving something that is not actually there. When we talk about LLMs hallucinating, we today[7] usually mean that LLMs make statements that do not correspond to reality or invent sources or facts that do not exist. The structural similarity here is the lack of correspondence between perception or statement and reality. Hicks et al. (2024) are right to emphasise that it is the representation of reality that is in question here.[8] However, even if we cannot attribute the intention to represent reality to LLMs, users are expected to ascribe it to the outputs produced because that is why they use the chatbot. They want to learn more about things and reality. Furthermore, it is unclear whether Hicks et al. (2024) are correct in their assertion that the metaphor implies that LLMs have a 'concern with the truth of their statements'. Even if humans who hallucinate are concerned with the truth of their perceptions, in that they believe them to be real, the metaphorical relation does not ascribe intentions to the hallucinating subject. Rather, it implies that a person who is hallucinating does not have complete control over their perceptions and thoughts.

---

[6] Both authors discuss the widespread use and the general importance of — often anthropomorphic — metaphors in the description of AI. Mitchell (2024) especially raises problems about the "AI-as-mind"-metaphor. Nehrlich (2024) presents a broad spectrum of different forms and functions of metaphors.
[7] For further differentiation see Gunkel & Coghlan (2025) and below.
[8] In their critique about Hicks et al.'s (2024) dismissal of hallucinating as a useful metaphor for LLMs, Gunkel and Coghlan (2025) do not talk about Hicks et al.' concern with the metaphor's implication that LLMs try to represent the empirical world. However, one could respond to this: LLMs might not aim to represent the real world, but they are designed to represent the world's knowledge about the real world and thereby implicitly also represent the world.

Here, insights which Gunkel and Coghlan (2025) present in their response to Hicks et al. (2024), are relevant. They emphasise that the associations with the metaphor "hallucination" have not ever been negative but rather affirmative towards creativity and inventiveness that can lie in hallucinations.[9] We should also note that this broader meaning originates from the complex history of the imagination, which has experienced long-term success since the Renaissance, when the imagination rose to become the central creative mental faculty in philosophy and literary theory. One of the main reasons for this success was the dichotomous concept of recognition, in which imagination plays a significant role in acquiring worldly knowledge. This culminated in Kant's concept of *Freies Spiel der Einbildungskraft* (free play of the imagination), which was intended to constitute art as a field in which the human capacity for freedom is most evident (see Liao & Gendler, 2020; Neuenfeld, 2004). Positive evaluations thus did not origin in the reliability of the imagination's representation of the empirical world but rather in its freedom from the constraints of a representational relation. Such freedom in relation to facts and reality does not seem welcome in LLMs. However, one could argue that a certain degree of freedom (or critical reflection) with regard to the training data could be beneficial in certain cases.

So, let us introduce a new dimension to the metaphor, albeit hypothetically. This comparison should aim to shed light on the relationship between human-produced and LLM-generated texts, given that the homogeneity of LLM output is what distinguishes it from human writing. The question, then, is: Can the hallucination metaphor adequately capture this? On the one hand, LLM-based chatbot texts seem similar to human-written texts in the same way that hallucinations resemble real perceptions; chatbot texts imitate average texts in terms of style, argument or tone. On the other hand, it appears that they do not. While hallucinations usually resemble real perceptions, chatbot texts consistently exhibit an average level of style which differs from individual human text production. However, it does not seem as if the use of the metaphor is intended to highlight this distinction either, because the distorted relationship with reality takes precedence in the perception of the metaphor, and the rhetorical qualities of chatbot texts are not considered, since hallucinations are generally associated with visual perception or imagination.

When we use "bullshit" in the sense introduced by Harry Frankfurt as a metaphor for LLM-based chatbots the *tertium* is the indifference between truth or falsehood of the AI-generated output. It has been argued that this comes with the advantage that we do not anthropomorphize the chatbot and do not ascribe the intention to represent reality correctly (Hicks et al., 2024; Fischer, 2025). However, in their respond to Hicks et al. (2024), Gunkel and Coghlan (2025) and Fisher (2025) argue that the term "bullshit" is associated with anthropomorphic meaning, too, and that it comes with disadvantages in terms of overcritical attitudes towards the new technology.[10]

But what about the rhetoric qualities of chatbots such as ChatGPT? Is it helpful to call them "bullshitters" in order to address these linguistic and rhetorical qualities? At first glance, it seems like an obvious choice because, in Harry Frankfurt's sense, bullshitters aim to manipulate their audience by conveying a message about themselves through talking bullshit about something else. It might be true that bullshit and manipulation as a means of bad rhetoric in the above-mentioned sense have something in common. However, the quality of averageness of style, argument and tone is, while being also rhetorically relevant, a distinct quality of LLM-produced texts. This quality is connected to probability and plausibility, which are achieved through alignment with the majority of human-produced texts and abstention from individual characteristics. The metaphor "bullshitting" does not seem to be helpful in this regard.

---

[9] Allen (2014) locates hallucinations entirely within the realm of the imagination, developing a positive description of the imagination that fits entirely within Kant's concept of the free play of the imagination; see also Maleki (2024).

[10] About anthropomorphic metaphors in talking about computers see Floridi and Nobre (2024).

Rudko and Bonab (2025) suggest a new distinction between what-BS (= what-bullshit) and how-BS (= how-bullshit) and argue that whereas what-BS is what Frankfurt has focused on, there is also a form of bullshit that has to do with the form of language and assertions. Rudko and Bonab (2025) use their new concept in order to explain why and how LLMs produce homogenous texts.

However, as the authors themselves admit, it is a completely newly built concept with loose connections to the original concept. They simply name the feature that LLMs produce homogenous texts "how-BS". This would only make sense if the production of homogeneous texts were related to indifference towards truth and falsehood. However, Rudko and Bonab (2025) do not make this claim. Therefore, we do not see why it makes sense to call the phenomenon "how-BS".

We do not deny the issue at hand, but if we use a term that has not been coined in another field or theory, it cannot function as a metaphor. It does not provide insights into the target object from another field because there is no other field. We therefore argue that Rudko and Bonab's (2025) description of "how-BS" could be more accurately conveyed by an existing metaphor with a scholarly tradition and differently located home base.

We propose the term "kitsch" for this purpose, since what Rudko and Bonab (2025) describe, e.g. about AI-generated images that look like indistinct boring mass-products, perfectly fits with aspects of "kitsch" and the scholarly debate about kitsch, which is associated with trivial literature or mass culture (Eco, 1986; Genz, 2011; Braungart, 2002).

Thus, "bullshit" and "hallucinating" work well for hinting at deviations from facts and sources. However, they do not perform well in naming and explaining rhetorical characteristics and problems with homogeneity of LLM-produced texts or images.

## 4. Talking about Kitsch: Defining Markers and Their LLM Parallels

As "hallucinations" and "bullshit" are insufficient to describe the problem of homogeneous texts produced by LLMs, we propose "kitsch" as a new metaphor. First, we will define and describe "kitsch" and its specific markers in order to prepare four arguments about why this is a fruitful metaphor for LLM-based chatbots and their strive for plausibility and the effect of homogeneity in texts and images.

Ever since the term emerged in the 19[th] century, the definition of "kitsch" has been much debated. (Friedrich, 2000; Kliche, 2001). Even though some scholars claim that "kitsch" cannot be defined at all (Braungart, 2002) or does not deserve a scholarly definition (Genz, 2011; Küpper, 2022), all concepts seem to agree that kitsch aims to appeal to the emotions (Braungart, 2022; Friedrich 2000; Dutton, 2003[11]). There is also mostly agreement that it is essential for kitsch to be easily accessible (Genz, 2011; Menninghaus, 2009). Most scholars also agree that kitsch has a tendency toward lighter entertainment and does not require arduous contemplation and rational reflection (Killy, 1962; pace Küpper, 2022[12]). This last description encompasses some of the most important and most debated characteristics that are used to distinguish it from art and (high) literature and set it apart from them. The history of kitsch theories is indispensably connected to the development of modern literary and art theories from the enlightenment and especially Kant (Schulte-Sasse, 1971) until postmodernism and today (Küpper, 2002[13]).

---

[11] Dutton (2003) presents a bunch of different definitions and qualifications and abstains from clearly defining "kitsch" himself.
[12] Küpper (2022) attempts to claim a special form of reflexivity for kitsch, namely a not purely sensual lower form of reflexivity.

Unlike art, which requires strenuous processes of understanding, kitsch immediately and superficially, i.e., deriving the emotionally moving factors from abstract patterns, evokes simple, strong emotions, making it convenient and enjoyable for many recipients.

However, this does not mean that kitsch and its reception are purely immediate and sensual, in the sense that there is no active, spontaneous, or reflective element (Küpper, 2022; Ackerknecht, 1950). We need to properly define this, as understanding the gradual differences of cognitive activity in kitsch and art is important for gaining a nuanced understanding of AI-generated products. To do this, we need to delve a little bit deeper into the history of aesthetics.

For, in scholarship, problems have arisen from the fact that kitsch and art have been defined according to modern aesthetics based on Baumgarten, Kant and subsequent Romanticism which assume a sharp dichotomy between passive perception and reflective spontaneous reason (Schmitt, 2008b). According to this framework, active cognitive components in an immediate and more sense-based form of perception, as in kitsch reception, cannot be thought of. Such a denial does not correspond with the phenomena. For example, if I do not somehow engage with the romantic love story of a telenovela — distinguishing between the characters, following the plot and empathising with the radiant heroes — I will not be moved by pity or enjoy the broadcasting.

In order to describe the difference between kitsch and other forms of literature and reception more accurately, we need to identify different levels of rational or intellectual involvement. This should correspond with the phenomena (compare the empirical studies by Sarkosh & Menninghaus, 2016)[14].

This option is provided in the Aristotelian conceptual epistemological framework for the reception of art and literature which distinguishes gradually between less rational receptions that stick to the surface of a story or performance on the one hand and those which get to the bottom of the story or representation and require further complex reflections on the other.

We will use this conceptual framework in order to distinguish between art and kitsch. We argue that the pleasure derived from kitsch does not stem from complex, provocative or disturbing insight, but rather from the familiarity and simplicity of a well-known plot-structure. These mean that there is cognitive effort required but this is rather low and finite (Braungart, 2002). The familiarity and simplicity of the plot structure and characters also make it easy to be moved (Hanich et al., 2014). We can conclude that both the easy attainability of the goal of being emotionally moved at all (Hanich et al., 2014)[15] and simple requirements for intellectual activity are central for the definition of kitsch.

This gives a hint at where to look for the structural reasons for these qualities. In order to find the answer, we need to ask: Under which circumstances and when do we find texts directly accessible, simply enjoyable, emotionally strong and entertaining? We argue that this is the case when they are plausible, have some probability and can be unlocked easily.

---

[13] Küpper (2002) and his argument against classical definitions of "kitsch" are rooted essentially in modern aesthetics.
[14] Sarkhosh and Menninghaus (2016) emphasize that, apart from the positive enjoyment, "trash films" are also received from an ironic stance which includes reflections on the quality of the movies.
[15] Hanich et al. (2024) explain by referring to the pleasure of being moved why people like to watch sad movies; however, these insights also point to another result, namely that the target emotion is rather abstract and not specific to the individual movie or its plot.

Probability has been a major concern for poets, literary theory as well as rhetoric ever since classical antiquity (Worthington, 1994). However, Aristotle calls for a distinction to be made between two types of probability that literary texts or texts in general can seek for: They can either be oriented towards the normal daily life and its patterns and stochastic probabilities (also as reflected in literature and art). If an author follows this trait, he or she will be likely to produce kitsch and clichéd characters and plots. Or, they can derive the probability about the characters' actions from their specific individual characters as Aristotle demands from the good poet (Aristotle. Poetics, ch. 9, Schmitt, 2008a).

For example: As Achilles is an exceptionally just character who always demands that everyone is treated fairly, it is plausible that he would act as Homer describes in the Iliad. He withdrew from battle and abandoned his comrades to near-total defeat because he had been treated unjustly by the commander, Agamemnon. This is despite the fact that he cared deeply for his friends and had always tried to protect them. Compared to the average behaviour of most people and what is probable in everyday life, his behaviour is completely unlikely and implausible. In order to understand his actions, listeners or readers need to follow closely and rationally recognize the specific individuality of Achilles, his outstanding sense of justice which led him astray in this particular situation because he could no longer pay attention to anything else than the injustice suffered. This is demanding and requires intellectual activity to understand. It cannot be automatically extracted as an abstract, general pattern. This is what we will call "literature" as opposed to "kitsch".

In contrast, if the action is designed directly from the plausibility of everyday behaviour and familiar, well-known patterns of action and simple structures, that reveal a turning point from fortune to misfortune as in many kitsch plots, there is no intellectually demanding task for the reader or recipient to fulfil. This plausibility appears to derive from its predictability, which in turn stems from its simple and trivial structures (that fail to do justice to the complexity of genuine human character). The recipient will be able to directly understand character and plot. This can be produced in an abstract manner by evoking emotions on the surface, using methods that work for many storylines.

For example, in the 20th century kitsch novelist Hedwig Courths-Mahler's romance novels, luminous yet socially disadvantaged heroes are invariably separated from their true loves by intrigue, ultimately achieving happiness and wealth through the power of love. These repetitive narrative patterns, intended to evoke 'great emotions', determine the plot rather than the characters' individual choices. This is what we will call a characteristic of kitsch.

The literary scholar Walther Killy (Killy, 1962) famously constructed a piece of literary text in order to exemplify what kitsch is like. He created a collage from many different works of (mostly) popular literature, quoting only descriptions and passages that not only fit a specific plot, but are generally applicable to all. He claimed that it is not possible to note that the pastiche is built out of several different stories, and that kitsch is structured in this way: by putting together plausible, well-known and approved elements and pieces in order to superficially evoke a certain atmosphere and emotional effect (Eco, 1989), and to realise an abstract structure and pattern in any material that will receive favourable resonance.

This also means that as kitsch can be produced just as easily by building on the same patterns and structures that have worked out well in other cases, mass production of kitsch is an immediate task and can be fulfilled by both humans and Generative AI. If one pattern works on humans then it is likely that others — humans or AI bots — will use the same pattern in order to get the same positive result and appreciation. Again, romances of Hedwig Courts-Mahler can illustrate this and its economic success (Küpper, 2012). The use of patterns in an iterative process of mass production is one of the most important similarities between kitsch and chatbot-generated texts.

Kitsch, thus, will have similarity and little diversity in structure and patterns of the created stories, because it relies on story lines and patterns that work well because of their structural affinity to everyday life. Therefore, we will see much homogeneity, average quality and similarity in kitsch.

However, on the other hand, kitsch seems to display reality other than it actually is and presents a more pleasant, more perfect world without the hardships that real life entails (Küpper, 2022): Kitsch often presents an unrealistic world with a happy ending guaranteed. It smooths out the rough edges of real life, so to speak, and presents a largely peaceful, pleasant world with the promise that, after intrigues, strokes of fate, and turmoil, a happy ending awaits. Many scholars have criticized that this is a bad illusionary mimesis (i.e., imitation) (Killy, 1962). However, kitsch does so in a predictable and expected manner and by fulfilling the audience's desire for illusion (Eco 1986). Everyone knows what they are getting into when they consume kitsch and agrees to the rules of it (Küpper, 2022).[16]

One property of kitsch is the polished style and (mostly) rather high stylistic level of language that kitsch adopts and cultivates (Eco, 1986; Küpper, 2022[17]) which often is just too much: an accumulation or pleonasm of words instead of a single word (Kliche 2001). Here, the same applies as for the plot structure and evoking of emotions: kitsch does not develop its linguistic vividness and stylistic quality from the individual plot or the individual subject of the representation, but rather derives it as an abstract essence from linguistically or artistically high-quality products and applies it uniformly everywhere in order to gain accessibility (Genz, 2011; Braungart, 2002) and pleasurable emotional experiences. A sublime or elevated style is added — sometimes in an inappropriate manner or in an exaggeratedly high style — like a gold overlay or sugar icing. As Umberto Eco put it: "In kitsch, however, the register shift has no cognitive function, but serves solely to intensify an emotional stimulus, and ultimately the 'insertion' [i.e., lyrical elements in order to produce some atmosphere] becomes the norm."[18]

In studies on kitsch, negative assessments and contrasts to "proper" literature are controversially discussed. Recent research tends to criticise critical assessments about kitsch (Küpper, 2022; Genz, 2011) and emphasize the social functions of kitsch or the value of diversity and a non-hierarchical concept of literature. Some argue that a general devaluation of kitsch is too undifferentiated and does not do justice to its complex aesthetic and social nature (Sarkhosh & Menninghaus, 2016; Hanich et al., 2014), thereby continuing earlier research that analysed the emotional function of kitsch which manages to evoke emotions by rather simple and general mechanisms (Eco, 1986; Killy, 1962; Broch, 1960).

In summary, it can be said that, given the breadth of its characteristics as described above, kitsch has the specific potential to serve as a metaphor for highlighting the homogeneity of AI-generated texts and the averaging process. We argue that kitsch and LLM-based chatbots share essential similarities in all the main areas, which we present in a condensed form in a short list below. The definition of kitsch, thus, provides us with the basis for concluding that there are four categories in which the metaphor of kitsch helps to deal with the homogeneity of AI-generated texts.

1) Both result from the abstraction of patterns.

---

[16] Küpper refers here to the fictionality contract, which is fundamental to approaches based on reception aesthetics (Nickel-Bacon et al., 2000).

[17] P. 131-166: This chapter describes the style of novelist Hedwig Courths-Maler, who became the very embodiment of kitsch writers in literary criticism.

[18] Eco (1986, 249): „Im Kitsch jedoch hat der Registerwechsel keine Erkenntnisfunktion, sondern dient ausschließlich der Verstärkung eines Gefühlsreizes, und letztlich wird die ‚Einlage' [i.e., lyrical elements in order to produce some atmosphere] zur Norm."

As we have seen, LLMs calculate the stochastic probability of word sequences and mimic the average of natural language texts, using algorithms to ensure they appear plausible to humans. This characteristic and its problematic effects can be described and elucidated using the "kitsch" metaphor. Kitsch draws on linguistic or visual patterns, probable storylines, character traits, and image patterns to produce easily accessible and plausible narratives or images. Kitsch reveals that this has certain advantages and disadvantages, and enables us to critically monitor, and if necessary, mitigate or remedy, them.

2) Both superficially simulate and evoke feelings and grandeur.

It has been argued in scholarship that LLMs or chatbots based on LLMs do not aim to state the truth. Rather, they incidentally create the impression that there is a claim for truth. This is because they mimic average texts, narratives, images, etc. LLMs support the probability of their output because their algorithms aim to create the impression of plausibility for humans. We argue that this can be described as the superficial addition of probability and plausibility which we have shown to be a specific characteristic of kitsch. There, the probability is not necessarily connected to the content produced. Furthermore, this addition is intended to evoke appreciation, i.e., positive emotions, in human users. When plausibility is added independently of the quality of the text or image, such abstract additions can be misused for manipulative purposes. We therefore need counterstrategies. Mechanisms like that can be studied in kitsch literature and art where strategies for moving the recipients emotionally are added to the stories or images as abstract features of the products. Analysis and interpretation in literary studies can be transferred to research in LLM and their societal impact in order to better understand mechanisms and remedies against it.

3) Both realize flawlessness and smooth or glossy language.

In literature and art, stylistic qualities that do not correspond to the genre, storyline or design of the image are easily detectable signals of kitsch. They are often exaggerated, use skewed linguistic images or figurative elements, and rely on an inappropriately high level of style. Kitsch embellishes average content with abstract qualities that are thought to be associated with beauty, pathos, or sympathy. Kitsch and LLMs both do not represent neither the phenomena of reality nor its essence truthfully but in an embellished version. This meets the criteria for a form of manipulation.

These qualities contribute to the perception that kitsch does not fulfil the criteria for good literature or art, since the constituent parts do not fit together and cannot be used to evaluate the content of a text or image. Analysing the relationship between the smooth and glossy styles of kitsch can therefore inform the analysis and improvement of LLMs. This can generate awareness of the disconnect between form and content, and highlight the need for a critical assessment of AI products. Furthermore, insights from kitsch studies can be used to implement algorithms that aim not only at plausibility, but also at truthfulness and the trustworthiness of the results presented (Agarwal et al., 2024).

4) Both reduce creativity and innovativeness and end in an averageness of language and structure.

Kitsch literature and art are not creative in themselves but profit from higher literature from which they abstract easily accessible features which are likely to be appreciated by the recipients and contribute to the emotional address (Eco, 1986). They derive the plausibility of stories and images from the orientation on everyday life and patterns of normal behaviour. Both mechanisms lead to mass products that share most characteristics and apply the same course of action or image design and the same formula to ever new — structurally similar — individual cases. The sameness of a multitude of products reduce creativity in authors and users or recipients and refer to a lack of innovative thinking.

As mass production kitsch floods the market and attracts the users' attention who get used to the average quality level of such products.

We face the same dangers with the ubiquity of texts and images from LLMs. There is a great need of empirical and theoretical studies about the societal impact of homogeneous and average outputs of LLMs.

Therefore, "kitsch" is best suited to highlight the characteristics and potential negative impact of LLMs producing average, homogeneous content.

## 5. Objections and Limitations, and Interdisciplinary Responses

This paper adopts a multidisciplinary approach by critically integrating literary theory on kitsch with contemporary research in AI, philosophy, and communication studies. This synthesis enables a comprehensive analysis of the structural homogeneity and rhetorical features of LLM-generated texts, enriching the discourse on both aesthetic and ethical dimensions of AI-driven content production.

A common objection concerns the primarily aesthetic origins of the kitsch concept, which might imply too narrow a focus on literary and artistic questions when applied to LLMs. We counter this by emphasising kitsch's longstanding political and social significance, rooted in industrial modernization and cultural conflicts (Kliche, 2001; Illing, 2018). For instance, Hermann Broch's (1960) ethical critique of kitsch highlights its role in fostering distorted perceptions of reality, demonstrating that kitsch transcends purely aesthetic boundaries. This broader framing validates the applicability of the metaphor to the societal effect of AI-generated content.

Nevertheless, important limitations remain. Future research must clarify how the "kitsch" metaphor can differentiate between varying degrees of truthfulness and distortion in AI-produced texts and images. Such refinements are necessary to better understand and address the nuances of representational fidelity and manipulation in LLM outputs. Furthermore, the focus on literary studies primarily addresses the transfer to textual output from LLMs. Although the rhetorical and aesthetic analyses are applicable to images produced by LLMs, expanding them to include other media, such as images and videos, is a task yet to be accomplished.

Interdisciplinary investigation is crucial, especially linking insights from kitsch studies on the production of embellished realities with findings from political communication research on echo chambers, filter bubbles, and social media effects. These processes often result in fragmented and personally tailored information environments, paralleling the homogenizing tendencies of AI-generated content.

On the technological front, progress in chatbot development may enhance linguistic and stylistic variety, either through algorithmic modifications to the probabilistic generation process or by introducing filtering layers before outputs reach users. However, since homogeneity is integral to probability-driven language modeling, computational improvements alone will not suffice, highlighting the need for complementary social, regulatory, and educational interventions.

## 6. Implications for Evolving AI Oversight and Ethics

The metaphor of kitsch provides valuable perspectives for understanding, analysing, and improving LLM-based chatbots such as ChatGPT. More broadly, it connects with ongoing reflections on the nature of literature and cultural production. When reality is sentimentally distorted and idealized in the manner of kitsch, this can foster addictive engagement, as users may internalise the softened, glorified depictions of life and identity offered by such content. Walter Benjamin's insights kitsch's dual capacity to both absorb reality into pleasant illusions and make these illusions immediately accessible are particularly instructive (Menninghaus, 2009). This issue is especially urgent in an online ecosystem increasingly saturated with AI-generated outputs. It raises foundational questions about the authenticity and appropriateness of emotional content, and invites scrutiny of the manipulative potentials inherent in such sentimentally charged communications. Literary studies, with their extensive bodies of work on the emotional and rhetorical mechanisms of kitsch — from Aristotle to contemporary scholarship — can deeply inform this inquiry.

There is significant potential for future interdisciplinary research and application. Building on our incisive critique of AI text homogeneity via the kitsch lens, subsequent studies should explore:

- Regulatory frameworks aimed at preserving stylistic diversity and guarding against the manipulative use of superficially plausible LLM outputs.
- Socio-political consequences of widespread homogenisation in AI-generated content.
- Governance models emphasising normative accountability and ethical oversight to balance innovation with societal well-being.

From a technical standpoint, efforts to foster creativity and counteract the homogenising effects of language models could draw on insights from kitsch studies regarding the aesthetic characteristics and social functions of homogenised mass products. There are potential solutions to the problem of homogenous and average output: 1. Generative adversarial networks (GANs) could be used. These are a type of deep learning architecture in which two neural networks compete with each other to generate better new data from a given training dataset (Goodfellow et al., 2014). This mutual evaluation and competition, especially when conceptually informed by insights from kitsch studies, could lead to output that is more closely connected to a specific task and context. This output would therefore abstain from the abstract relationship between general plausibility and specific content. 2. Further methods are currently being developed to enable LLMs to evaluate each other, with the aim of verifying and refining the results (Guo et al., 2025). One of them could identify and detect only superficially convincing results and try to alter them. 3. ThoughtSculpt: This new method uses an LLM-powered thought evaluator to provide feedback on candidate partial outputs. A component of the thought generator then produces potential solutions. The aim is to achieve better, verified results (Chi et al., 2025). Like RAG (Retrieval Augmented Generation), this method can also assist with fact-checking. Further research is needed to evaluate if or how these methods may also mitigate the problems associated with non-pertinent plausibility or whether they tend to exacerbate the problem.

Educational initiatives are likewise essential, promoting public awareness of AI-generated content's rhetorical strategies and their cognitive and emotional impacts. Such literacy programs, informed by kitsch studies, could empower users to engage more critically and resist manipulative persuasion.

Finally, by combining the metaphor of kitsch with political communication concepts like echo chambers and filter bubbles, there is impetus for developing Responsible Foundation Models—AI systems designed not only for reliability and factuality but also for cultural and rhetorical pluralism.

## 7. Conclusions

We have demonstrated that "kitsch" can serve as a useful metaphor for LLMs. This helps us to better understand and address the insufficiently researched tendency of AI to produce homogeneous and average texts and images. Accordingly, "kitsch" can be a useful metaphor in debates about AI and LLMs. This is particularly pertinent when the objective is not merely to describe factual inaccuracies ("hallucinations", "bullshit"), but to encapsulate a fundamental characteristic of AI: the creation of average, relatively non-variant content, such as texts and images. "Hallucinations" and "bullshit" are both unsuitable for addressing this characteristic and its impact on communication, the reception of AI-generated texts and images, and human image and text production itself.

"Kitsch" has been proven helpful due to an essential characteristic identified or – often controversially – discussed by literary and cultural studies and philosophy in the 20th and 21st centuries. Among them, the fact that kitsch uses average and probable linguistic patterns and storylines is of great importance, as these correspond to what we usually expect to be possible. Kitsch is based on rhetoric probability because it aims to persuade the audience to enjoy and be moved by the similar, easily apprehensible stories and characters, which are designed to be superficially appealing. This means that kitsch is always easily accessible, requiring no intensive intellectual activity, and allowing one to simply be entertained while simultaneously believing that one is reading or watching literature or art. Kitsch studies have already discussed aesthetic, social, educational and communicative problems that arise from kitsch's strive for easy access and immediate emotional effect. As there is no direct link between plausibility and content, recipients also cannot verify the plausibility of a text or image, nor are they encouraged to do so. For this reason, kitsch can be exploited for manipulative purposes. Understanding this is important for analysing and countering potential manipulation by LLM-based chatbots.

The problem of reduced creativity and innovativeness among human content creators is of great importance when considering these effects. Further research into mitigating these effects by modifying existing algorithms could be supported by kitsch studies and studies in literature and art in general.

The same applies to the strong emotions evoked by kitsch literature and art, which can divert attention away from any negative consequences. It is becoming increasingly apparent that LLM-based chatbots have the potential to manipulate users. This makes insights from kitsch studies extremely valuable.

The tendency of kitsch literature and art to soften reality and present an unrealistic vision of an ideal world can also provide important insights that we can apply elsewhere. For example, insights from the study of echo chambers or filter bubbles in political communication can be combined and revised to enhance the development of Responsible Foundation Models.

Due to its pattern-oriented production method, kitsch can be mass-produced, with the same economically successful narrative and linguistic elements repeated in each instance. The impact of kitsch mass production on the literary and art scene has been studied by kitsch scholars, with findings that can easily be transferred to LLM studies.

Building upon the critical examination of the structural homogeneity inherent in LLM outputs through the lens of kitsch, future research should explicitly engage in urgent questions of regulation, societal and political consequences, as well as governance frameworks, thereby connecting these considerations to ongoing debates on normative frames and accountability in the development and deployment of responsible AI systems.